# The wave-particle duality of the qudit quantum space and the quantum wave gates

Zixuan Hu and Sabre Kais*

*Department of Chemistry, Department of Physics, and Purdue Quantum Science and Engineering Institute, Purdue University, West Lafayette, IN 47907, United States*
*Email:* kais@purdue.edu

Abstract: We propose three core ideas: 1. the wave-particle duality of the qudit quantum space; 2. the classification of all elementary quantum gates by ordered pairs of qudit functionals; 3. a new type of quantum gates called the "quantum wave gates". We first study the quantum functionals whose relation to the quantum states is analogous to that between the momentum and position wavefunctions in fundamental quantum physics: a Fourier transform and an entropic uncertainty principle can be defined between the dual representations. The quantum functionals are not just mathematical constructs but have clear physical meanings and quantum circuit realizations. Connecting the partition interpretation of the qudit functionals to the effects of quantum gates we classify all elementary quantum gates by ordered pairs of qudit functionals. By generalizing the qudit functionals to quantum functionals, the new type of "quantum wave gates" are discovered as quantum versions of the conventional quantum gates.

## 1. Introduction

Quantum information and quantum computation has received enormous interests in recent years [1-6] with the potential to revolutionize information processing and computing technologies [7-12]. In particular, numerous quantum algorithms have been developed for solving a variety of problems in the last few decades: e.g. the phase estimation algorithm [13], Shor's factorization algorithm [14], the Harrow-Hassidim-Lloyd algorithm for linear systems [15], the hybrid classical-quantum algorithms [16, 17], the quantum machine learning algorithms [18-20], and quantum algorithms for open quantum dynamics [21-25].

Despite the diverse selection of quantum algorithm designs, there is the need for a theory [26-30] to systematically understand the structures and behaviors of the quantum circuits used to implement the quantum algorithms. Qudits are the fundamental building blocks of quantum information and quantum computation – usually the 2-level qubits are used, but recently there are growing interests in generalizing to higher $d$-level qudit systems [31-35]. In a quantum computer, the creation, manipulation and measurement of quantum states are all based on qudits. Therefore, a deeper understanding of the qudit quantum space is essential to developing a systematic theory for designing efficient quantum circuits.

In this study of the qudit quantum space – which can be specialized to the conventional qubit space – we propose three core ideas: 1. the wave-particle duality of the qudit quantum space; 2. the classification of all elementary quantum gates by ordered pairs of qudit functionals; 3. a new type of quantum gates called the "quantum wave gates".

We first propose a new "quantum functional space" that is generated by the basis set consisting of the "qudit functionals". By the mathematical theory of Pontryagin duality [36, 37], the qudit functionals are the duals of the basis states of the usual quantum state space, and therefore the relation between the quantum functionals and quantum states is analogous to the relation between the momentum and position wavefunctions commonly known in fundamental quantum physics: a Fourier transform and an entropic uncertainty principle can be defined between the dual representations. Here we propose the quantum functionals and quantum states are respectively the wave representation and particle representation of the same qudit quantum object, and thus there is also a wave-particle duality for the qudit quantum space. Indeed, we find the basis of the quantum functionals, i.e. the qudit functionals, can be interpreted as planewaves in the state-representation where the wavenumber **k** characterizes the fluctuations of the wavefunction in the state space.

We then show the wave-like quantum functionals are not just mathematical constructs but physical objects, as the qudit functionals can be physically interpreted as partitions of the qudit quantum state space, and a physical realization on the quantum circuit can be constructed for the quantum functional associated with any arbitrary quantum state. Relating the partition interpretation of the qudit functionals to the effects of quantum gates, we propose that all elementary quantum gates – i.e. 1-qudit unitaries and 2-qudit controlled-unitaries – can be classified into classes each defined by an ordered pair of qudit functionals: in particular, the control functional defines to which subspace the unitary is applied and the target functional defines the rule for how the unitary should be applied. This classification identifies quantum gates with the qudit functionals, which means quantum gates can be understood as wave-like objects of the qudit quantum space. Then by generalizing the qudit functionals to quantum functionals, we discover an entirely new type of "quantum wave gates" that can be interpreted as quantum versions of the conventional quantum gates. We then design the quantum circuits that realize two examples of the quantum wave gates. The classification of all conventional quantum gates by the qudit functionals means the quantum gate properties can be indexed by the wavenumber **k**, which may allow the systematic characterization of quantum gates. The new type of quantum wave gates will be a useful addition to the toolbox of quantum circuit design.

## 2. The wave-particle duality of the qudit quantum space.

### 2.1 The quantum functional space.

For an *n*-qudit system with *d* levels on each qudit, an arbitrary quantum state can be written as a linear combination $|\psi\rangle = \sum_{j=0}^{d^n-1} a_j |j\rangle$ of the basis states $|j\rangle$ that are essentially strings of qudit values: e.g. any 2-qutrit ($d=3$) state vector has $3^2 = 9$ basis states corresponding to 9 qutrit-value-strings: $|00\rangle \sim (0,0)$, $|01\rangle \sim (0,1)$, $|02\rangle \sim (0,2)$, $|10\rangle \sim (1,0)$, …, $|22\rangle \sim (2,2)$. Without affecting the main results, for the following discussion we assume *d* is a prime number *p*, and the case of *d* not prime is discussed in the Supplementary Information (SI). Then mathematically the complete collection of the qudit-value-strings (i.e. basis states) of any $n$-qudit quantum space forms a linear space $Q$ over the field $\mathbb{Z}_p$, which means any vector $\mathbf{q}$ in $Q$ is:

$$\mathbf{q} = (q_1, q_2, ..., q_n) = \sum_{j=1}^{n} q_j \mathbf{q}^{(j)} \qquad q_j = 0, 1, 2, ..., p-1 \qquad (1)$$

where $\mathbf{q}^{(j)}$ is the standard unit vector $(0, ..., 0, q_j = 1, 0, ..., 0)$; the scalars $q_j$'s take integer values from 0 to $p-1$ (i.e. values from $\mathbb{Z}_p$); the addition is digit-wise addition modulo $p$. Using the 2-qutrit example again, any vector $\mathbf{q}$ in $Q$ can be expressed as a linear combination of the 2 basis vectors (1,0) and (0,1): $\mathbf{q} = (q_1, q_2) = q_1(1,0) + q_2(0,1)$, where the coefficients $q_1$ and $q_2$ take integer values 0, 1 or 2 (i.e. values from $\mathbb{Z}_3$). By the mathematical theory of linear spaces, all the linear functionals defined on $Q$ form another linear space $Q^*$ that is the dual space of $Q$. For reasons that will become obvious later, we label a vector in $Q^*$ by $\mathbf{k}$, which can be written as:

$$\mathbf{k} = (k_1, k_2, ..., k_n) = k_1 q_1 \oplus k_2 q_2 \oplus ... \oplus k_n q_n \qquad k_j = 0, 1, 2, ..., p-1 \qquad (2)$$

where the scalars $k_j$'s take integer values from 0 to $p-1$; $\oplus$ is addition modulo $p$. Clearly, $\mathbf{k}$ is a linear combination of the basis functionals (basis vectors of the functional space $Q^*$): $f^{(j)}(\mathbf{q}) = q_j$. Some examples of the linear functionals defined on the 2-qutrit space are $(0,1) = q_2$, $(1,2) = q_1 \oplus 2q_2$, $(2,1) = 2q_1 \oplus q_2$, and there are totally $3^2 = 9$ of these linear functionals in $Q^*$ – matching the number of qudit-value-strings in $Q$. Here the linear functionals in $Q^*$ are essentially functionals of qudit values and in the following we will call them the "qudit functionals".

Now we notice that the vectors $\mathbf{q}$ in $Q$ are basis states that generate the usual quantum state space such that any quantum state is written as $|\psi\rangle = \sum_{\mathbf{q} \in Q} a_\mathbf{q} |\mathbf{q}\rangle$, where the summation goes over all $\mathbf{q}$ in $Q$ and $a_\mathbf{q}$ are complex numbers. Then naturally, the dual vectors of $\mathbf{q}$, i.e. the qudit functionals can also be basis states, or more appropriately "basis functionals", to generate the "quantum functional space" where an arbitrary quantum functional is written as:

$$|\phi\rangle = \sum_{\mathbf{k} \in Q^*} b_\mathbf{k} |\mathbf{k}\rangle, \qquad b_\mathbf{k} \text{ is a complex number that } \sum_{\mathbf{k} \in Q^*} |b_\mathbf{k}|^2 = 1 \qquad (3)$$

Here $|\phi\rangle = \sum_{\mathbf{k} \in Q^*} b_\mathbf{k} |\mathbf{k}\rangle$ has the same quantum interpretation as $|\psi\rangle = \sum_{\mathbf{q} \in Q} a_\mathbf{q} |\mathbf{q}\rangle$ such that $b_\mathbf{k}$ represents the probability amplitude of the corresponding qudit functional $|\mathbf{k}\rangle$. In fact, as the dual space $Q^*$ has all the mathematical properties of $Q$ itself, the quantum functional space generated by $\mathbf{k}$ in $Q^*$ also has all the mathematical properties of the usual quantum state space generated by **q** in $Q$: this means there are also quantum entities like superposition and entanglement in the quantum functional space. Again using the 2-qutrit case for example, the quantum functional $|\phi_1\rangle = (b_0|0\rangle + b_1|1\rangle) \otimes |2\rangle$ is a "superposition functional" for which the $k_1$ entry of $|\mathbf{k}\rangle = |k_1 k_2\rangle = |k_1\rangle \otimes |k_2\rangle = (k_1, k_2) = k_1 q_1 \oplus k_2 q_2$ is in the superposition of 0 and 1, and thus the total functional is in the superposition of $(0,2) = 2q_2$ and $(1,2) = q_1 \oplus 2q_2$, with the probability amplitudes $b_0$ and $b_1$ respectively. Similarly, the quantum functional $|\phi_2\rangle = b_0|12\rangle + b_1|21\rangle$ is an "entangled functional" such that the $k_1$ and $k_2$ entries are correlated in a quantum way with no classical equivalent.

## 2.2 The Fourier transform and the wave-particle duality.

So far we have shown the quantum functional space is generated by the qudit functionals $|\mathbf{k}\rangle$, and the latter are dual vectors of the basis states $|\mathbf{q}\rangle$ of the usual quantum state space. It turns out the relation between **q** and **k** is analogous to the relation between the position variable **x** and the momentum variable **k** in fundamental quantum physics. In particular, **x** and **k** are also dual vectors and quantum wavefunctions can be expressed in either the position representation $\psi(\mathbf{x})$ or the momentum representation $\phi(\mathbf{k})$. Mathematically, both the **x**-**k** duality and the **q**-**k** duality are examples of the Pontryagin duality [36, 37] that guarantees a Fourier transform between wavefunctions expressed in the dual representations. For the **x**-**k** pair we have the Fourier transform:

$$\psi(\mathbf{x}) = \int_{\mathbf{k}-space} \phi(\mathbf{k}) e^{2\pi i \mathbf{k} \cdot \mathbf{x}} d\mathbf{k} \qquad \phi(\mathbf{k}) = \int_{\mathbf{x}-space} \psi(\mathbf{x}) e^{-2\pi i \mathbf{k} \cdot \mathbf{x}} d\mathbf{x} \qquad (4)$$

where $\psi(\mathbf{x})$ is the probability amplitude evaluated at $|\mathbf{x}\rangle$ in the position space, and $\phi(\mathbf{k})$ is the probability amplitude evaluated at $|\mathbf{k}\rangle$ in the momentum space. Similarly for the **q**-**k** pair we have the Fourier transform:

$$\psi(\mathbf{q}) = \frac{1}{\sqrt{p^n}} \sum_{\mathbf{k}} \phi(\mathbf{k}) e^{2\pi i \mathbf{k} \cdot \mathbf{q}/p} \qquad \phi(\mathbf{k}) = \frac{1}{\sqrt{p^n}} \sum_{\mathbf{q}} \psi(\mathbf{q}) e^{-2\pi i \mathbf{k} \cdot \mathbf{q}/p} \qquad (5)$$

where $p$ is the dimension of one qudit, $n$ is the number of qudits, $\mathbf{k} \cdot \mathbf{q} = k_1 q_1 \oplus k_2 q_2 \oplus ... \oplus k_n q_n$ ($\oplus$ is addition modulo $p$), $\psi(\mathbf{q})$ is the probability amplitude evaluated at $|\mathbf{q}\rangle$ in the $Q$ space, and $\phi(\mathbf{k})$ is the probability amplitude evaluated at $|\mathbf{k}\rangle$ in the $Q^*$ space: clearly $\psi(\mathbf{q}) = a_\mathbf{q}$ corresponds to $|\psi\rangle = \sum_{\mathbf{q} \in Q} a_\mathbf{q} |\mathbf{q}\rangle$ and $\phi(\mathbf{k}) = b_\mathbf{k}$ corresponds to $|\phi\rangle = \sum_{\mathbf{k} \in Q^*} b_\mathbf{k} |\mathbf{k}\rangle$. A graphical illustration of the **q**-**k** duality and the Fourier transform between $\psi(\mathbf{q})$ and $\phi(\mathbf{k})$ is shown in Figure 1 with a 2-qutrit example.

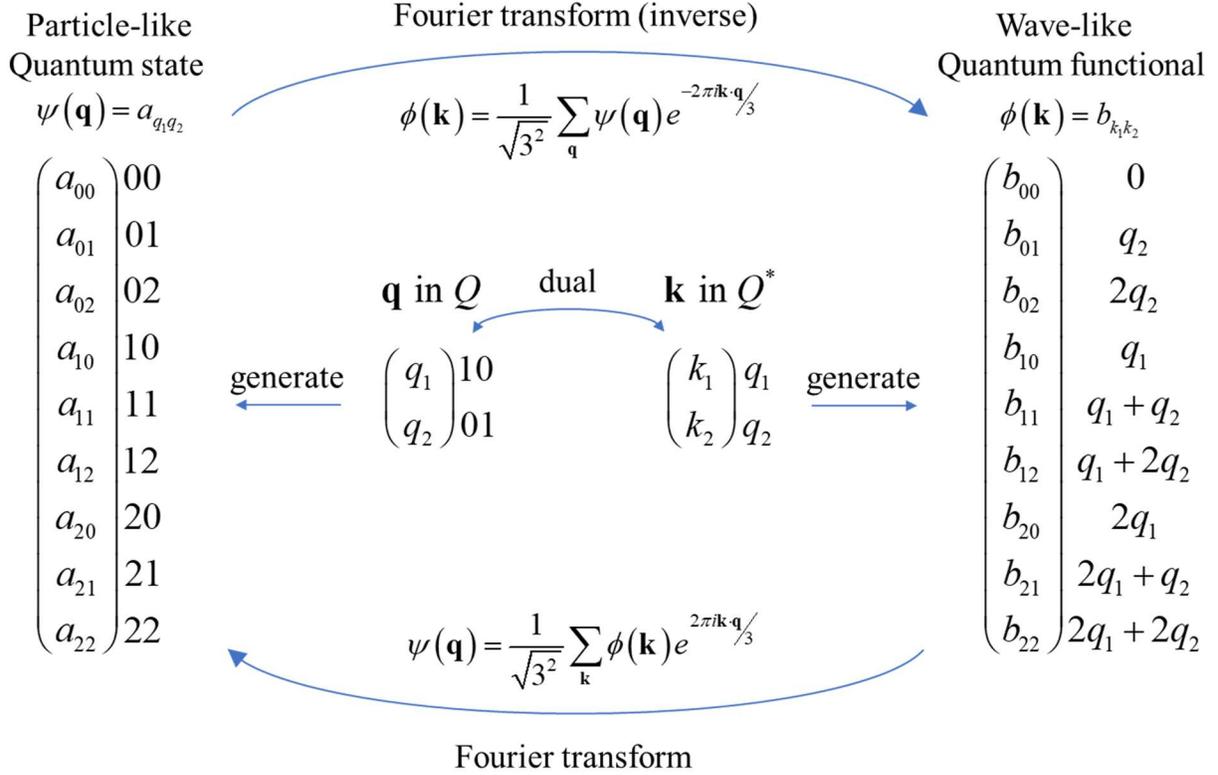

Figure 1. Graphical illustration of the **q**-**k** duality and the Fourier transform between the wavefunctions $\psi(\mathbf{q})$ and $\phi(\mathbf{k})$, using a 2-qutrit system as an example. The **q** and **k** vectors in the middle are duals of each other. The particle-like quantum state generated by **q** and the wave-like quantum functional generated by **k** are related by a Fourier transform.

The similarity between the Fourier transforms in Equations (4) and (5) clearly shows the equivalence between the **q**-**k** duality and the **x**-**k** duality. In particular, the **x**-**k** duality is commonly interpreted as the wave-particle duality where $\psi(\mathbf{x})$ is the particle representation that describes the spread of the quantum state in the position space, and $\phi(\mathbf{k})$ is the wave representation that describes the spread of the quantum state in the momentum space. Here we propose the 1st core idea of the study:

> **Idea 1:** The **q**-**k** duality can be interpreted as the wave-particle duality of the qudit quantum space where $\psi(\mathbf{q})$ is the particle representation that describes the spread of the quantum state in the basis state **q** space, and $\phi(\mathbf{k})$ is the wave representation that describes the spread of the quantum state in the qudit functional **k** space.

To better see the wave character of $\phi(\mathbf{k})$, consider the basis functional $|\mathbf{k}\rangle$: its wavefunction in the $\mathbf{k}$-representation is $\phi(\mathbf{k}) = \delta_{\mathbf{k},\mathbf{j}}$ ($\delta_{\mathbf{k},\mathbf{j}}$ is the Kronecker delta), so $|\mathbf{k}\rangle$ in the $\mathbf{q}$-representation is obtained by the Fourier transform in Equation (5):

$$\langle \mathbf{q} | \mathbf{k} \rangle = \psi(\mathbf{q}) = \frac{1}{\sqrt{p^n}} \sum_{\mathbf{j}} \delta_{\mathbf{k},\mathbf{j}} e^{2\pi i \mathbf{j} \cdot \mathbf{q}/p} = \frac{1}{\sqrt{p^n}} e^{2\pi i \mathbf{k} \cdot \mathbf{q}/p} \tag{6}$$

where $\langle \mathbf{q} | \mathbf{k} \rangle$ has the form of the planewave similar to the usual $e^{2\pi i \mathbf{k} \cdot \mathbf{x}}$, and $\mathbf{k} \cdot \mathbf{q} = k_1 q_1 \oplus k_2 q_2 \oplus ... \oplus k_n q_n$ can be interpreted as the dot product between the wavenumber vector and the position vector. Now we see that labeling the qudit functionals by the symbol $\mathbf{k}$ is appropriate as it indeed plays the role of the wavenumber vector that specifies the fluctuations of the wavefunction in the $\mathbf{q}$ space: each component $k_j$ is the wavenumber on the direction of $q_j$. In a similar manner we find $\langle \mathbf{k} | \mathbf{q} \rangle = \frac{1}{\sqrt{p^n}} e^{-2\pi i \mathbf{k} \cdot \mathbf{q}/p}$ and thus the Fourier transform in Equation (5) can be understood as taking the inner product of the wavefunction with the basis of the dual space:

$$\psi(\mathbf{q}) = \langle \mathbf{q} | \phi \rangle_{\mathbf{k}} = \sum_{\mathbf{k}} \left( \frac{1}{\sqrt{p^n}} e^{-2\pi i \mathbf{k} \cdot \mathbf{q}/p} \right)^* \phi(\mathbf{k}) \qquad \phi(\mathbf{k}) = \langle \mathbf{k} | \psi \rangle_{\mathbf{q}} = \sum_{\mathbf{q}} \left( \frac{1}{\sqrt{p^n}} e^{2\pi i \mathbf{k} \cdot \mathbf{q}/p} \right)^* \psi(\mathbf{q})$$

$$\langle f(\mathbf{k}) | g(\mathbf{k}) \rangle_{\mathbf{k}} \equiv \sum_{\mathbf{k}} f^*(\mathbf{k}) \cdot g(\mathbf{k}) \qquad \langle f(\mathbf{q}) | g(\mathbf{q}) \rangle_{\mathbf{q}} \equiv \sum_{\mathbf{q}} f^*(\mathbf{q}) \cdot g(\mathbf{q}) \tag{7}$$

where the inner product in each representation is defined below the corresponding Fourier transform. It follows that any quantum functional $|\phi\rangle = \sum_{\mathbf{k}} b_{\mathbf{k}} |\mathbf{k}\rangle$ is indeed the linear combination of the planewaves $\frac{1}{\sqrt{p^n}} e^{2\pi i \mathbf{k} \cdot \mathbf{q}/p}$ and thus $|\phi\rangle$ is a wave-like quantum object.

### 2.3 The physical reality of the qudit functionals and quantum functionals.

The qudit functionals $|\mathbf{k}\rangle$, the quantum functionals $|\phi\rangle = \sum_{\mathbf{k}} b_{\mathbf{k}} |\mathbf{k}\rangle$, and their roles in the wave-particle duality of the qudit quantum space have been established by the mathematical theory of Pontryagin duality. In this section we show that the qudit functionals and quantum functionals are not only mathematical constructs but physical objects with clear physical meanings and realizations.

**2.3.1 The realization of the qudit functionals as partitions of the quantum state space.** As shown in Equation (2), the qudit functional $\mathbf{k}$ is defined as the linear functional of the basis state $\mathbf{q}$, and therefore any $\mathbf{k}$ can be interpreted as a "partition" of the quantum state space. Using the 2-qutrit example again, say a qudit functional $\mathbf{k}_1 = (0,1) = q_2$, then it defines the "qudit conditions"

"$q_2=0$", "$q_2=1$", and "$q_2=2$", which respectively specify the subspace spanned by $\{|00\rangle,|10\rangle,|20\rangle\}$, $\{|01\rangle,|11\rangle,|21\rangle\}$, and $\{|02\rangle,|12\rangle,|22\rangle\}$: this is a partition of the total quantum state space into three equal subspaces. Similarly, $\mathbf{k}_2=(2,1)=2q_1\oplus q_2$ defines the qudit conditions "$2q_1\oplus q_2=0$", "$2q_1\oplus q_2=1$", and "$2q_1\oplus q_2=2$", which respectively specify the subspaces spanned by $\{|00\rangle,|11\rangle,|22\rangle\}$, $\{|01\rangle,|20\rangle,|12\rangle\}$, and $\{|10\rangle,|21\rangle,|02\rangle\}$: this is a different partition of the total quantum state space into three different equal subspaces. It is simple to use the partition of $\mathbf{k}_1=(0,1)=q_2$ on a quantum circuit: just use the 2nd qutrit $q_2$ as the control, and then apply to the target the gate $U_0$ when $q_2=|0\rangle$, $U_1$ when $q_2=|1\rangle$, and $U_2$ when $q_2=|2\rangle$. We see that a controlled-gate can be understood as having its effect conditional on the value of the functional held by the control qudit, and in the $\mathbf{k}_1=(0,1)=q_2$ case we do not need to do anything to $q_2$ before using it as the control. In the $\mathbf{k}_2=(2,1)=2q_1\oplus q_2$ case however, we need to first create the functional on some qudit before using it as the control. This can be done by applying the following gate:

$$CU_{1\to 2}(2)=\begin{pmatrix} \mathbf{U}_0 & & \\ & \mathbf{U}_1 & \\ & & \mathbf{U}_2 \end{pmatrix}, \quad \mathbf{U}_0=\mathbf{I}, \quad \mathbf{U}_1=\begin{pmatrix} 0 & 1 & 0 \\ 0 & 0 & 1 \\ 1 & 0 & 0 \end{pmatrix}, \quad \mathbf{U}_2=\begin{pmatrix} 0 & 0 & 1 \\ 1 & 0 & 0 \\ 0 & 1 & 0 \end{pmatrix} \quad (8)$$

where $CU_{1\to 2}(2)$ is a controlled-gate using $q_1$ as the control and $q_2$ as the target, and $\mathbf{U}_1$ and $\mathbf{U}_2$ are the "translations" that advance the value of $q_2$ by 2 and 1 (4 modulo 3) respectively. Clearly, the $CU_{1\to 2}(2)$ in Equation (8) adds $2q_1$ to $q_2$ modulo 3, and thus $\mathbf{k}_2=(2,1)=2q_1\oplus q_2$ is now stored on $q_2$: i.e. using $q_2$ as the control for any subsequent controlled-gate will now use the partition defined by $\mathbf{k}_2=(2,1)=2q_1\oplus q_2$ to determine its effects on the target qudit. Now $CU_{1\to 2}(2)$ can be easily generalized to $n$-qudit with $p$ levels for each qudit:

$$CU_{1\to m}(k_1)=\begin{pmatrix} \mathbf{U}_0 & & & \\ & \mathbf{U}_1 & & \\ & & \ddots & \\ & & & \mathbf{U}_{p-1} \end{pmatrix}, \quad \mathbf{U}_j=\mathbf{T}_m(k_1 j) \quad j=0,1,...,p-1 \quad (9)$$

where $\mathbf{T}_m(k_1 j)$ is the translation that advances the value of $q_m$ by the product $k_1 j$, and clearly $CU_{1\to m}(k_1)$ adds $k_1 q_1$ to $q_m$ and creates $\mathbf{k}_3=(k_1,0,...,0,k_m=1,0,...,0)=k_1 q_1\oplus q_m$ on $q_m$ if $q_m$ is in its original functional. With some additional quantum gate algebra we can construct the gate $CU_{l\to m}(k_l)$ that adds any $k_l$ multiples of any $q_l$ to any $q_m$ and thus any qudit functional $\mathbf{k}$ can be created on any qudit.
7

**2.3.2 The realization of the quantum functionals on the quantum circuit.** Having seen the physical interpretation and realization of the qudit functionals, we now construct the quantum circuit to realize any quantum functionals $|\phi\rangle = \sum_{\mathbf{k}} b_{\mathbf{k}} |\mathbf{k}\rangle$:

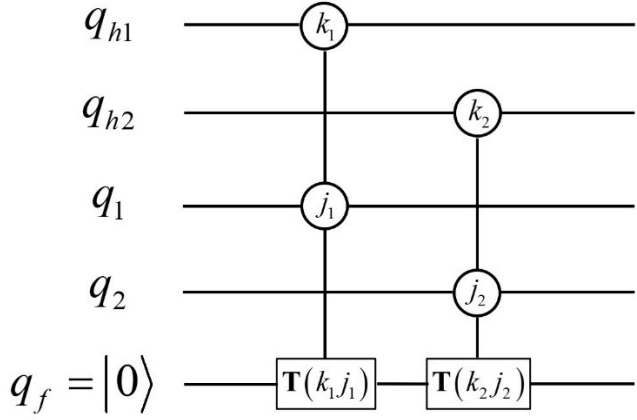

Figure 2. The quantum circuit for realizing any arbitrary 2-qudit quantum functional. In the two 3-qudit controlled-gates (qudit versions of the Toffoli gates), the handle qudits $q_{h1}$ and $q_{h2}$ control the values $k_1$ and $k_2$ that specify respectively how many multiples of $q_1$ and $q_2$ are added to the functional holder $q_f$ by the translations $\mathbf{T}(k_1 j_1)$ and $\mathbf{T}(k_2 j_2)$. Any arbitrary 2-qudit quantum functional $|\phi\rangle = \sum_{k_1,k_2=0}^{p-1} b_{k_1 k_2} |k_1 k_2\rangle$ defined on the space of $q_1$ and $q_2$ can be created on $q_f$ by initializing $|\phi\rangle$ as a 2-qudit state on the handle qudits $q_{h1}$ and $q_{h2}$ and then applying the circuit.

In Figure 2 we have constructed the quantum circuit to realize any quantum functional that can be defined on the 2-qudit system of $q_1$ and $q_2$. There are two 3-qudit controlled-gates that are qudit generalizations of the Toffoli gates. The 1$^{\text{st}}$ gate uses $q_1$ and the handle qudit (the qudit used to handle the quantum functionals) $q_{h1}$ as the controls and the functional holder qudit $q_f$ as the target. The circle with $k_1$ inside indicates $q_{h1}$ controls the value of $k_1$ in the translation $\mathbf{T}(k_1 j_1)$, and the circle with $j_1$ inside indicates $q_1$ controls the value of $j_1$ in $\mathbf{T}(k_1 j_1)$. The same mechanism applies to how the 2$^{\text{nd}}$ gate is controlled by $q_{h2}$ and $q_2$. $q_f$ is initialized to $|0\rangle$ so its value does not contribute to the functional created on it. Now if the handle qudits $q_{h1}$ and $q_{h2}$ take definite values $|k_1\rangle$ and $|k_2\rangle$, then the two 3-qudit controlled-gates are just implementations of the $CU_{l \to m}(k_l)$ gate discussed above for adding $k_l$ multiples of $q_l$ to $q_m$, and thus the circuit creates the qudit functional $|\mathbf{k}\rangle = |k_1 k_2\rangle = k_1 q_1 \oplus k_2 q_2$ on $q_f$: this means any controlled-gate using $q_f$ as the control and either $q_1$ or $q_2$ as the target will use the partition of $|\mathbf{k}\rangle$ to determine its effect on the 2-qudit system of $q_1$ and $q_2$. However, if the handle qudits are set to be quantum states like $q_{h1} = \sum_{j=0}^{p-1} a_j |j\rangle$ and $q_{h2} = \sum_{j=0}^{p-1} b_j |j\rangle$, then the circuit will create the quantum functional $|\phi\rangle = \left(\sum_{k_1=0}^{p-1} a_{k_1} |k_1\rangle\right) \otimes \left(\sum_{k_2=0}^{p-1} b_{k_2} |k_2\rangle\right)$ on $q_f$: which means any controlled-gate using $q_f$ as the control and either $q_1$ or $q_2$ as the target will use the partition of $|\mathbf{k}\rangle = |k_1 k_2\rangle = k_1 q_1 \oplus k_2 q_2$ with the probability of $|a_{k_1}|^2 \cdot |b_{k_2}|^2$ – this is a typical quantum interpretation. Indeed, one can verify that any arbitrary 2-qudit quantum functional $|\phi\rangle = \sum_{k_1,k_2=0}^{p-1} b_{k_1 k_2} |k_1 k_2\rangle$ can be created on $q_f$ by first initializing $|\phi\rangle$ as a 2-qudit state on the handle qudits $q_{h1}$ and $q_{h2}$, and then applying the circuit in Figure 2. It is also straightforward to extend the 2-qudit case to any *n*-qudit case. This circuit design can also be specialized to conventional qubit systems as will be shown in Section 3.2.

The wave-particle duality means any qudit quantum object has dual characters of particle and wave, and therefore any arbitrary particle-like quantum state $|\psi\rangle = \sum_{\mathbf{q}} a_{\mathbf{q}} |\mathbf{q}\rangle$ in the **q**-representation is inherently associated with a wave-like quantum functional $|\phi\rangle = \sum_{\mathbf{k}} b_{\mathbf{k}} |\mathbf{k}\rangle$ in the **k**-representation. Now given $|\psi\rangle = \sum_{\mathbf{q}} a_{\mathbf{q}} |\mathbf{q}\rangle$, we can easily perform the Fourier transform in Equation (5) to obtain the associated $|\phi\rangle = \sum_{\mathbf{k}} b_{\mathbf{k}} |\mathbf{k}\rangle$, and then realize $|\phi\rangle$ with the quantum circuit in Figure 2 (as generalized to *n*-qudit). This means the quantum functional inherently associated with any quantum state can be realized and utilized with a simple quantum circuit setup.

## 3. The classification of elementary quantum gates and the quantum wave gates.

Having seen the interpretation of the qudit functionals as partitions of the quantum space and circuit realizations of both the qudit functionals and the quantum functionals, next we introduce the 2nd and 3rd core ideas of this study:

> **Idea 2:** All elementary quantum gates – i.e. 1-qudit unitaries and 2-qudit controlled-unitaries – can be classified into classes each defined by an ordered pair of qudit functionals.

> **Idea 3:** The quantum functionals as generated by the basis of qudit functionals lead to the new type of "quantum wave gates", which are quantum versions of the conventional quantum gates.

## 3.1 The classification of elementary quantum gates by qudit functionals.

For clarify and simplicity, in the following discussions we use the conventional 2-level qubits as examples, and only mention the general qudit case when it provides additional insight and knowledge unique to higher-level qudit systems. Any quantum circuit can be decomposed into a sequence of elementary quantum gates that are either 1-qubit unitaries or 2-qubit controlled-unitaries. Consider the 2-qubit controlled-unitaries applied to a 2-qubit state $|\psi\rangle = (a_1, a_2, a_3, a_4)^T$:

$$CU_{1\to 2}|\psi\rangle = \begin{pmatrix} 1 & 0 & 0 & 0 \\ 0 & 1 & 0 & 0 \\ 0 & 0 & u_1 & u_2^* \\ 0 & 0 & u_2 & -u_1^* \end{pmatrix} \begin{pmatrix} a_1 \\ a_2 \\ a_3 \\ a_4 \end{pmatrix} \begin{matrix} |00\rangle \\ |01\rangle \\ |10\rangle \\ |11\rangle \end{matrix} = \begin{pmatrix} a_1 \\ a_2 \\ u_1 a_3 + u_2^* a_4 \\ u_2 a_3 - u_1^* a_4 \end{pmatrix} \begin{matrix} |00\rangle \\ |01\rangle \\ |10\rangle \\ |11\rangle \end{matrix}$$

(10)

$$CU_{2\to 1}|\psi\rangle = \begin{pmatrix} 1 & 0 & 0 & 0 \\ 0 & u_1 & 0 & u_2^* \\ 0 & 0 & 1 & 0 \\ 0 & u_2 & 0 & -u_1^* \end{pmatrix} \begin{pmatrix} a_1 \\ a_2 \\ a_3 \\ a_4 \end{pmatrix} \begin{matrix} |00\rangle \\ |01\rangle \\ |10\rangle \\ |11\rangle \end{matrix} = \begin{pmatrix} a_1 \\ u_1 a_2 + u_2^* a_4 \\ a_3 \\ u_2 a_2 - u_1^* a_4 \end{pmatrix} \begin{matrix} |00\rangle \\ |01\rangle \\ |10\rangle \\ |11\rangle \end{matrix}$$

In the top row of Equation (10) $CU_{1\to 2}$ applies $U = \begin{pmatrix} u_1 & u_2^* \\ u_2 & -u_1^* \end{pmatrix}$ to $q_2$ if $q_1 = |1\rangle$, and does nothing if $q_1 = |0\rangle$: this means $CU_{1\to 2}$ uses the qubit functional $f_1 = q_1$ to partition the space into $\{|00\rangle, |01\rangle\}$ where $U$ is not applied and $\{|10\rangle, |11\rangle\}$ where $U$ is applied. In addition, when $U$ is applied to modify $a_3$ and $a_4$, it uses the qubit functional $f_2 = q_2$ to further partition the space into $\{|10\rangle\}$ and $\{|11\rangle\}$ such that $a_3$ is considered as the $|0\rangle$ term while $a_4$ as the $|1\rangle$ term in effectively $\begin{pmatrix} u_1 & u_2^* \\ u_2 & -u_1^* \end{pmatrix}\begin{pmatrix} a_3 \\ a_4 \end{pmatrix} = \begin{pmatrix} u_1 a_3 + u_2^* a_4 \\ u_2 a_3 - u_1^* a_4 \end{pmatrix}$. Now similarly for $CU_{2\to 1}$ in the bottom row of Equation (10), the qubit functional $f_2 = q_2$ is used to partition the space into $\{|00\rangle, |10\rangle\}$ where $U$ is not applied and $\{|01\rangle, |11\rangle\}$ where $U$ is applied. In addition, when $U$ is applied to $a_2$ and $a_4$, it uses the qubit functional $f_1 = q_1$ to further partition the space into $\{|01\rangle\}$ and $\{|11\rangle\}$ such that $a_2$ is considered as the $|0\rangle$ term while $a_4$ as the $|1\rangle$ term in effectively $\begin{pmatrix} u_1 & u_2^* \\ u_2 & -u_1^* \end{pmatrix}\begin{pmatrix} a_2 \\ a_4 \end{pmatrix} = \begin{pmatrix} u_1 a_2 + u_2^* a_4 \\ u_2 a_2 - u_1^* a_4 \end{pmatrix}$. By these examples we see that the effect of a 2-qubit controlled-unitary can be specified by an ordered pair of qubit functionals: the control qubit functional specifies how to partition the space such that we know where to apply the unitary, and the target qubit functional specifies how to further partition the space such that we know which term is considered as $|0\rangle$ and which as $|1\rangle$ for the unitary. Note that the ordered pair of qubit functionals does not fix the values of $u_1$ and $u_2$ in $U$, so what it really defines is a class of 2-qubit unitaries with similar effects for modifying the state vector. In the above cases, $CU_{1\to 2}$ belongs to the class of $(q_1, q_2)$ (where the 1st term is the control functional and the 2nd term is the target functional) and $CU_{2\to 1}$ belongs to the class of $(q_2, q_1)$. In general, two 2-qubit unitaries belonging to the same class apply $U$ in the same subspace defined by the control partition and apply $U$ according to the same rule defined by the target partition – consequently these two gates will have similar properties and can be continuously morphed into each other. On the other hand, like $CU_{1\to 2}$ and $CU_{2\to 1}$ here, two 2-qubit unitaries belonging to different classes will have distinct properties and cannot be continuously morphed into each other. Different classes do not share any gate except when $U$ is the identity.

In Section 2.1 the space $Q^*$ formed by all qudit functionals is the dual of the space $Q$ formed by all qudit-strings. For the current 2-qubit space, $Q^*$ has $2^2 = 4$ elements: 0, $q_1$, $q_2$, and $q_1 \oplus q_2$. Now by specializing the gate in Equation (8) to the qubit case or by studying the truth table of the CNOT gate, we can add the value of $q_1$ to $q_2$ and thus set the functional on $q_2$ to be $f_2 = q_1 \oplus q_2$ by applying $CX_{1\to 2}$ to $|\psi\rangle$, and then this functional will be available for use in any subsequent 2-qubit controlled-unitaries:

$$CU_{1\to 2}CX_{1\to 2}|\psi\rangle = \begin{pmatrix} 1 & 0 & 0 & 0 \\ 0 & 1 & 0 & 0 \\ 0 & 0 & u_1 & u_2^* \\ 0 & 0 & u_2 & -u_1^* \end{pmatrix}\begin{pmatrix} 1 & 0 & 0 & 0 \\ 0 & 1 & 0 & 0 \\ 0 & 0 & 0 & 1 \\ 0 & 0 & 1 & 0 \end{pmatrix}\begin{pmatrix} a_1 \\ a_2 \\ a_3 \\ a_4 \end{pmatrix}\begin{matrix}|00\rangle \\ |01\rangle \\ |10\rangle \\ |11\rangle\end{matrix} = \begin{pmatrix} a_1 \\ a_2 \\ u_1 a_4 + u_2^* a_3 \\ u_2 a_4 - u_1^* a_3 \end{pmatrix}\begin{matrix}|00\rangle \\ |01\rangle \\ |10\rangle \\ |11\rangle\end{matrix}$$

(11)

$$CU_{2\to 1}CX_{1\to 2}|\psi\rangle = \begin{pmatrix} 1 & 0 & 0 & 0 \\ 0 & u_1 & 0 & u_2^* \\ 0 & 0 & 1 & 0 \\ 0 & u_2 & 0 & -u_1^* \end{pmatrix}\begin{pmatrix} 1 & 0 & 0 & 0 \\ 0 & 1 & 0 & 0 \\ 0 & 0 & 0 & 1 \\ 0 & 0 & 1 & 0 \end{pmatrix}\begin{pmatrix} a_1 \\ a_2 \\ a_3 \\ a_4 \end{pmatrix}\begin{matrix}|00\rangle \\ |01\rangle \\ |10\rangle \\ |11\rangle\end{matrix} = \begin{pmatrix} a_1 \\ u_1 a_2 + u_2^* a_3 \\ a_4 \\ u_2 a_2 - u_1^* a_3 \end{pmatrix}\begin{matrix}|00\rangle \\ |01\rangle \\ |10\rangle \\ |11\rangle\end{matrix}$$

In the top row of Equation (11) $CU_{1\to 2}$ is applied after $CX_{1\to 2}$, so its ordered pair of functionals or class is now $(q_1, q_1 \oplus q_2)$. Compared with applying $CU_{1\to 2}$ only in Equation (10), the control functional is the same $q_1$, but the target functional is changed from $q_2$ to $q_1 \oplus q_2$. We see indeed in the top row of Equation (11) $U$ is applied to the same subspace $\{|10\rangle, |11\rangle\}$ where $q_1 = |1\rangle$, but its rule on this subspace is defined by $a_4$ as the $|0\rangle$ term while $a_3$ as the $|1\rangle$ term in effectively

$$\begin{pmatrix} u_1 & u_2^* \\ u_2 & -u_1^* \end{pmatrix} \begin{pmatrix} a_4 \\ a_3 \end{pmatrix} = \begin{pmatrix} u_1 a_4 + u_2^* a_3 \\ u_2 a_4 - u_1^* a_3 \end{pmatrix}$$ – this is the result of having the functional $f_2 = q_1 \oplus q_2$ on $q_2$ such that $a_3 \sim |10\rangle \sim 1 \oplus 0 = 1$ is now considered as the $|1\rangle$ term and $a_4 \sim |11\rangle \sim 1 \oplus 1 = 0$ is now considered as the $|0\rangle$ term – therefore the class $(q_1, q_1 \oplus q_2)$ behaves distinctly from $(q_1, q_2)$. Now similarly for $CU_{2 \to 1}$ in the bottom row of Equation (11), its class is now $(q_1 \oplus q_2, q_1)$, so compared with applying $CU_{2 \to 1}$ only in Equation (10), the control functional is changed from $q_2$ to $q_1 \oplus q_2$ while the target functional is the same $q_1$. We see indeed in the bottom row of Equation (11) $U$ is applied to the subspace $\{|01\rangle, |10\rangle\}$ where $q_1 \oplus q_2 = |1\rangle$, and its rule on this subspace is defined by $f_1 = q_1$ such that $a_2 \sim |01\rangle \sim 0$ is considered as the $|0\rangle$ term and $a_3 \sim |10\rangle \sim 1$ is considered as the $|1\rangle$ term – therefore the class $(q_1 \oplus q_2, q_1)$ behaves distinctly from $(q_2, q_1)$. This example also shows the control functional is more important than the target functional because even the target functional is the same $f_1 = q_1$ and thus the rule for the subspace is the same, the subspace itself is different with a different control functional $f_2 = q_1 \oplus q_2$ – consequently the difference between $CU_{2 \to 1} CX_{1 \to 2}$ and $CU_{2 \to 1}$ is greater than that between $CU_{1 \to 2} CX_{1 \to 2}$ and $CU_{1 \to 2}$.

In the previous example, applying $CX_{1 \to 2}$ adds the value of $q_1$ to $q_2$ and set $f_2 = q_1 \oplus q_2$. The two qubits then hold the functionals $\{f_1 = q_1, f_2 = q_1 \oplus q_2\}$ that form a "qubit functional configuration" (abbreviated as the "QFC" below; the symbol $\{\,,\,\}$ here and in the following represents a QFC) as detailed in Ref. [26]. For $n$ qubits, any QFC can only hold $n$ current functionals out of the total possible number of $2^n$, so any gate class realizable on the current QFC has to choose 2 functionals out of the $n$ current functionals to be the control and the target. For the above 2-qubit example, the original state $|\psi\rangle$ has the QFC of $\{f_1 = q_1, f_2 = q_2\}$, so there are 2 realizable gate classes: $(q_1, q_2)$ and $(q_2, q_1)$ as in Equation (10). On the other hand, applying $CX_{1 \to 2}$ to $|\psi\rangle$ makes the QFC of $\{f_1 = q_1, f_2 = q_1 \oplus q_2\}$, so there are 2 different realizable gate classes: $(q_1, q_1 \oplus q_2)$ and $(q_1 \oplus q_2, q_1)$ as in Equation (11). How many different 2-qubit controlled-unitary classes are possible for an $n$-qubit system? To realize a certain gate class, we need to first prepare the control functional and the target functional to be available in the current QFC, so the limitations on the QFC itself (derived in Ref. [26]) limits the gate classes. In particular, a valid QFC must not have any two functionals being the same form (e.g. $\{f_1 = q_1, f_2 = q_1\}$) or any functional being the 0 functional (e.g. $\{f_1 = q_1, f_2 = 0\}$), because 1. it will cause the quantum state to lose dimensions; 2. it is impossible to achieve QFC's that violate these rules by applying any sequence of CNOT gates to a starting QFC that satisfies these rules (see Ref. [26] for details). So for any gate class, we can choose 2 different functionals out of the $2^n - 1$ functionals (excluding the 0 functional), and

13permute them to have $P(2^n-1,2)=(2^n-1)\cdot(2^n-2)$ possible gate classes. It is straightforward to generalize this result to an *n*-qudit system: $P(p^n-1,2)=(p^n-1)\cdot(p^n-2)$.

The previous paragraph calculates the possible number of 2-qubit controlled-unitary classes. Now for completeness, we also define the 1-qubit unitary classes as those having the control functional being the 0 functional, in e.g. $(0,q_2)$ or $(0,q_1\oplus q_2)$. As discussed above, these classes are not realizable on any QFC because the 0 functional is invalid in any QFC. However, here the ordered pair such as $(0,q_2)$ is not meant to be realized as a 2-qubit controlled-unitary, but instead interpreted as the following: the control qubit will always be $|0\rangle$, so if we apply *U* when the control is $|1\rangle$ then it is never applied and the gate is effectively the identity; if we apply *U* when the control is $|0\rangle$ then it is always applied so the gate is effectively a 1-qubit unitary applied to the target qubit according to the rule specified by the target functional – therefore we can define the classes with the control functional being 0 as the 1-qubit unitary classes. This way our classification of the elementary quantum gates is complete as it includes both the 1-qubit unitaries and 2-qubit controlled-unitaries.

All discussions in this section can be easily generalized to qudits and lead to the conclusion that all elementary quantum gates – i.e. 1-qudit unitaries and 2-qudit controlled-unitaries – can be classified into classes each defined by an ordered pair of qudit functionals. This is a significant idea because it identifies quantum gates with the duals of the qudit-strings as detailed in Section 2.1: this means quantum gates can be understood as wave-like objects, i.e. planewaves of the qudit quantum space. Recall that in Section 2.1 the qudit functionals have been used as basis functionals to generate the new quantum space called the quantum functional space. Now applying the same idea to the quantum gate classes by having quantum functionals in the control and/or the target, we propose an entirely new type of quantum gates that we call the "quantum wave gates".

**3.2 The quantum wave gate classes as defined by ordered pairs of quantum functionals.**

In the previous section all elementary quantum gates can be assigned to classes defined by ordered pairs of qudit functionals. Then naturally, when we go from the qudit functionals to quantum functionals as discussed in Section 2.1, we will now have quantum versions of the elementary quantum gates. Considering the qudit functionals are the classical outcomes used as the basis of the quantum functionals, the conventional quantum gate classes defined by qudit functionals may be called the "classical quantum gates", while the new type of quantum gate classes defined by quantum functionals may be called the "quantum quantum gates". However, these names are potentially confusing and awkward, and therefore we have chosen to call the conventional "classical quantum gates" as "conventional quantum gates", and call the new type of "quantum quantum gates" as "quantum wave gates", which appropriately emphasizes their origin in the wave-like quantum functionals.

Again consider the 2-qubit state $|\psi\rangle = (a_1, a_2, a_3, a_4)^T$ in the starting QFC of $\{f_1 = q_1, f_2 = q_2\}$, by adding two handle qubits $q_{h_1}$ and $q_{h_2}$, one holder qubit $q_3$, and specializing the quantum circuit design in Figure 2 to qubits, we can create a quantum functional on $f_2$ by the circuit in Figure 3:

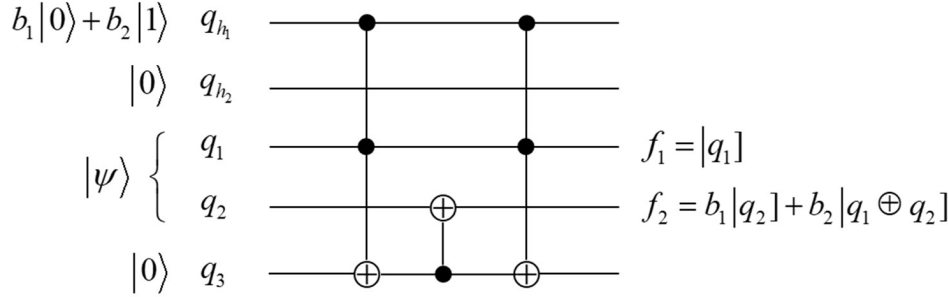

Figure 3. Quantum circuit realization of a quantum wave gate (superposition) on the 2-qubit space formed by $q_1$ and $q_2$. The symbol $|\cdot]$ emphasizes the basis such as $|q_1]$ and $|q_1 \oplus q_2]$ are qubit functionals. The 1st Toffoli gate $CCX_{h_1, 1 \to 3}$ uses the handle qubit $q_{h_1}$ to create the quantum functional $b_1|0] + b_2|q_1]$ on $f_3$, which is then added to $f_2$ by $CX_{3 \to 2}$, and finally $f_3$ is reset to $|0]$ by the 2nd Toffoli gate. Applying $CU_{1 \to 2}$ after this circuit gives the quantum wave gate class of $(|q_1], b_1|q_2] + b_2|q_1 \oplus q_2])$; applying $CU_{2 \to 1}$ gives $(b_1|q_2] + b_2|q_1 \oplus q_2], |q_1])$.

What happens after each gate in Figure 3 is explained by Equation (12):

$$\{f_1 = |q_1], f_2 = |q_2]\}(f_3 = |0]) \xrightarrow{CCX_{h_1, 1 \to 3}} \{f_1 = |q_1], f_2 = |q_2]\}(f_3 = b_1|0] + b_2|q_1])$$
$$\xrightarrow{CX_{3 \to 2}} \{f_1 = |q_1], f_2 = b_1|q_2] + b_2|q_1 \oplus q_2]\}(f_3 = b_1|0] + b_2|q_1]) \quad (12)$$
$$\xrightarrow{CCX_{h_1, 1 \to 3}} \{f_1 = |q_1], f_2 = b_1|q_2] + b_2|q_1 \oplus q_2]\}(f_3 = |0])$$

In Equation (12) the symbol $|\cdot]$ emphasizes the basis such as $|0]$ and $|q_1 \oplus q_2]$ are qubit functionals: in the $|\mathbf{k}\rangle = |k_1 k_2\rangle = |k_1 q_1 \oplus k_2 q_2]$ representation in Section 2.1, $|0]$ is $|00\rangle$ and $|q_1 \oplus q_2]$ is $|11\rangle$. Now the 1st Toffoli gate $CCX_{h_1, 1 \to 3}$ uses the quantum states on the handle qubits (in this particular case only $q_{h_1}$ is used) to set the quantum functional $f_3 = b_1|0] + b_2|q_1]$ on $q_3$. Then after $CX_{3 \to 2}$, $f_2$ becomes $b_1|q_2] + b_2|q_1 \oplus q_2]$. Finally the 2nd Toffoli gate $CCX_{h_1, 1 \to 3}$ resets $f_3$ to $|0]$, and the final QFC is $\{f_1 = |q_1], f_2 = b_1|q_2] + b_2|q_1 \oplus q_2]\}$ with $f_2$ being a quantum functional. Now applying $CU_{1 \to 2}$ will result in the gate class of $(|q_1], b_1|q_2] + b_2|q_1 \oplus q_2])$ on the 2-qubit space formed by $q_1$ and $q_2$, and applying $CU_{2 \to 1}$ will result in $(b_1|q_2] + b_2|q_1 \oplus q_2], |q_1])$ – both these classes are "quantum" in the sense that they involve quantum superposition functionals in the control or the target.

Because the quantum functional space is a fully functional quantum space having the same mathematical structure as the quantum state space (see Section 2.1), we can also design "entangled gate classes" that involve entangled functionals. Again start with $|\psi\rangle = (a_1, a_2, a_3, a_4)^T$ in the starting QFC of $\{f_1 = q_1, f_2 = q_2\}$, add two handle qubits $q_{h_1}$ and $q_{h_2}$, and one holder qubit $q_3$, then consider the circuit in Figure 4:

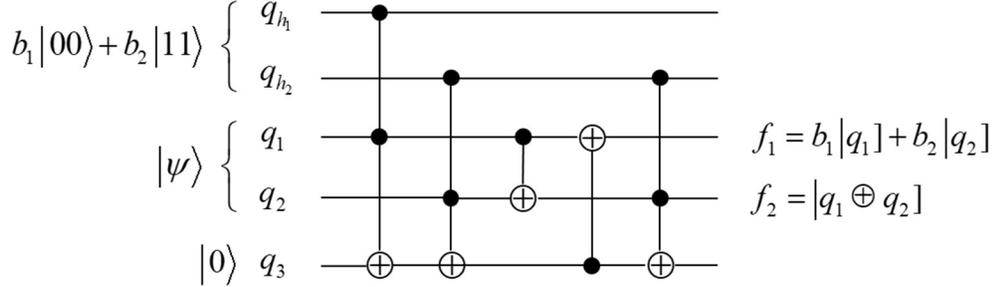

Figure 4. Quantum circuit realization of a quantum wave gate (entanglement) on the 2-qubit space formed by $q_1$ and $q_2$. The first two Toffoli gates $CCX_{h_1, 1 \to 3}$ and $CCX_{h_2, 2 \to 3}$ use the handle qubits $q_{h_1}$ and $q_{h_2}$ to create the entangled functional $b_1|0] + b_2|q_1 \oplus q_2]$ on $f_3$, which is then added to $f_1$ by $CX_{3 \to 1}$, and finally $f_3$ is reset to $|0]$ by the last Toffoli gate. The additional gate $CX_{1 \to 2}$ ensures all QFC outcomes are valid. Applying $CU_{1 \to 2}$ after this circuit gives the entangled gate class of $(b_1|q_1] + b_2|q_2], |q_1 \oplus q_2])$; applying $CU_{2 \to 1}$ gives $(|q_1 \oplus q_2], b_1|q_1] + b_2|q_2])$.

What happens after each gate in Figure 4 is explained by Equation (13):

$$\begin{aligned}
\{f_1 = |q_1], f_2 = |q_2]\}(f_3 = |0]) &\xrightarrow{CCX_{h_1,1 \to 3} \text{ and } CCX_{h_2,2 \to 3}} \{f_1 = |q_1], f_2 = |q_2]\}(f_3 = b_1|0] + b_2|q_1 \oplus q_2]) \\
&\xrightarrow{CX_{1 \to 2}} \{f_1 = |q_1], f_2 = |q_1 \oplus q_2]\}(f_3 = b_1|0] + b_2|q_1 \oplus q_2]) \\
&\xrightarrow{CX_{3 \to 1}} \{f_1 = b_1|q_1] + b_2|q_2], f_2 = |q_1 \oplus q_2]\}(f_3 = b_1|0] + b_2|q_1 \oplus q_2]) \\
&\xrightarrow{CCX_{h_2,2 \to 3}} \{f_1 = b_1|q_1] + b_2|q_2], f_2 = |q_1 \oplus q_2]\}(f_3 = |0])
\end{aligned} \quad (13)$$

In Equation (13) by using an entangled state $b_1|00\rangle + b_2|11\rangle$ on the handle qubits $q_{h_1}$ and $q_{h_2}$, and two Toffoli gates, we create the entangled functional $b_1|0] + b_2|q_1 \oplus q_2]$ on $f_3$. We then create the entangled functional $b_1|q_1] + b_2|q_2]$ on $f_1$ by $CX_{3 \to 1}$. The additional gate $CX_{1 \to 2}$ ensures all outcomes of the quantum functional are valid QFC's as explained in Section 3.1, and the final Toffoli gate resets $f_3$ to $|0]$. The final QFC is $\{f_1 = b_1|q_1] + b_2|q_2], f_2 = |q_1 \oplus q_2]\}$, and thus now applying $CU_{1 \to 2}$ will result in the gate class of $(b_1|q_1] + b_2|q_2], |q_1 \oplus q_2])$ on the 2-qubit space formed by $q_1$ and $q_2$, and applying $CU_{2 \to 1}$ will result in $(|q_1 \oplus q_2], b_1|q_1] + b_2|q_2])$ – where the functional $b_1|q_1] + b_2|q_2]$ is $b_1|10\rangle + b_2|01\rangle$ in the $|\mathbf{k}\rangle = |k_1 k_2\rangle = |k_1 q_1 \oplus k_2 q_2]$ representation such that it is a Bell-like quantum functional.

16With the two examples in Figure 3 and Figure 4 we have demonstrated the quantum circuit realization of an entirely new type of quantum gate classes – the quantum wave gate classes – which are nonetheless logical extensions of the conventional quantum gate classes by going from qubit functionals to quantum functionals. The discovery of the quantum wave gate classes can thus be considered as an application of the wave-particle duality theory established in Sections 2.1 through 2.3. Being quantum versions of the conventional quantum gates, the quantum wave gate classes can lead to entirely new ways for designing new quantum algorithms: this is an idea for a future study.

All results in this section can be easily generalized to a general $n$-qudit system without much effort. One notable difference for a $p$-level qudit with $p > 2$ is the space $Q^*$ formed by all qudit functionals will be much greater than the qubit case: in particular, as the wavenumber $k$ can take more values than 0 and 1, both the quantum gate and the quantum wave gate classes are more abundant. In the Supplementary Information (SI), we study the wavenumber $k$ as a physical observable and present an entropic uncertainty principle between the $k$ and $q$ observables. As the qudit functionals are indexed by $k$, the $k$ value can be used to systematically characterize physical properties of both the conventional quantum gates and the quantum wave gates: this is another idea for a future study.

## 4. Conclusions

In this study, we have proposed three core ideas: 1. the wave-particle duality of the qudit quantum space; 2. the classification of all elementary quantum gates by ordered pairs of qudit functionals; 3. the quantum functionals lead to the new type of quantum wave gates, which are quantum versions of the conventional quantum gates.

The wave-particle duality as demonstrated by the wave-like quantum functionals is a fundamental physical reality of the qudit quantum space. The quantum functionals are quantum objects generated by the basis consisting of qudit functionals, which are the duals of the usual basis states of qudit quantum states. The relation between the quantum states and quantum functionals is analogous to the relation between the position and momentum wavefunctions in fundamental quantum physics. In particular, there is a Fourier transform and an entropic uncertainty principle between the quantum functionals and quantum states.

Physically, the qudit functionals can be interpreted as partitions of the state space, and a quantum circuit has been designed to realize and utilize the quantum functionals. By relating this partition interpretation of the qudit functionals to the effects of quantum gates, we can classify all elementary quantum gates by ordered pairs of qudit functionals, which identify conventional quantum gates with planewaves of the qudit space. When generalizing the qudit functionals to quantum functionals, we naturally discover a new type of quantum gates called the quantum wave gates that can be interpreted as quantum versions of the conventional quantum gates.

The wave-particle duality of the qudit quantum space opens entirely new perspectives for studying quantum states and quantum gates. The classification of all elementary gates by qudit functionals and the discovery of the quantum wave gates show the theory is not only abstract but has important potential applications in quantum circuit design.

## 5. Acknowledgements

ZH thanks Peng Zhou for mathematical discussions. ZH and SK acknowledge funding by the U.S. Department of Energy (Office of Basic Energy Sciences) under Award No. DE-SC0019215, the National Science Foundation under award number 1955907, and CCI Phase I: NSF Center for Quantum Dynamics on Modular Quantum Devices (CQD-MQD).

## 6. Supplementary Information is available after the References.

# Supplementary Information: The wave-particle duality of the qudit quantum space and the quantum wave gates

Zixuan Hu and Sabre Kais*

*Department of Chemistry, Department of Physics, and Purdue Quantum Science and Engineering Institute, Purdue University, West Lafayette, IN 47907, United States*
*Email:* kais@purdue.edu

### S1. The wavenumber k as a physical observable.

The wave-particle duality means any qudit quantum object has dual characters of particle and wave, and therefore any quantum state in the **q**-representation can also be evaluated by an observable $\hat{\mathbf{k}}$ to get the expectation value of **k** inherently associated to the state. Focusing on a single dimension $\hat{k}$ first, in the *k*-representation it is clearly a diagonal matrix:

$$K_k = diag(0,1,...,p-1) \qquad \text{S(1)}$$

and its form in the *q*-representation can be obtained by the Fourier-transform such that:

$$K_q = FK_kF^{-1}, \quad F = \frac{1}{\sqrt{p}}\begin{pmatrix} 1 & 1 & \cdots & 1 \\ 1 & \omega_p^1 & \cdots & \omega_p^{p-1} \\ \vdots & \vdots & \ddots & \vdots \\ 1 & \omega_p^{p-1} & \cdots & \omega_p^{(p-1)^2} \end{pmatrix}, \quad \omega_p^p = 1 \qquad \text{S(2)}$$

where $F$ is the matrix form of the single-dimension Fourier transform from *k* to *q* as specialized from Equation (5) in the main text, and $\omega_p$ is the *p*-th root of 1. Clearly $K_q = FK_kF^{-1}$ gives the matrix form of the observable $\hat{k}$ in any single dimension of $\hat{\mathbf{k}} = (\hat{k}_1, \hat{k}_2, ..., \hat{k}_n)$, and now given any quantum state $|\psi\rangle = \sum_{\mathbf{q}} a_{\mathbf{q}}|\mathbf{q}\rangle = \sum_{\mathbf{q}} a_{\mathbf{q}}|q_1\rangle \otimes |q_2\rangle \otimes ... \otimes |q_n\rangle$, just apply $K_q$ to a qudit $q_j$ to evaluate $\langle \hat{k}_j \rangle$, and $\langle \hat{\mathbf{k}} \rangle = (\langle \hat{k}_1 \rangle, \langle \hat{k}_2 \rangle, ..., \langle \hat{k}_n \rangle)$ will be evaluated after obtaining each $\langle \hat{k}_j \rangle$. Here we have implicitly used the fact that the commutator $[\hat{k}_j, \hat{k}_h] = 0$ so all the $\hat{k}_j$ can be evaluated simultaneously. Now if we consider the observable $\hat{q}_j$ that evaluates the expectation value of $q_j$ of a quantum state, then in the *q*-representation the matrix form of $\hat{q}_j$ is $Q_q = diag(0,1,...,p-1)$, and we have:

$$[\hat{q}_j, \hat{k}_j] = [Q_q, K_q] \neq 0 \qquad \text{S(3)}$$

which is similar to the fact that $\left[\hat{x}_j, \hat{k}_j\right] \neq 0$ in the usual position-momentum duality, and thus we can conclude that $\hat{q}_j$ and $\hat{k}_j$ are incompatible variables that cannot be measured simultaneously. Indeed, by the Fourier transform in Equation (5) in the main text, if a wavefunction $\phi(\mathbf{k})$ is concentrated on a single $\mathbf{k}$ then its $\mathbf{q}$-representation $\psi(\mathbf{q})$ is uniformly distributed over $\mathbf{q}$; similarly if a wavefunction $\psi(\mathbf{q})$ is concentrated on a single $\mathbf{q}$ then its $\mathbf{k}$-representation $\phi(\mathbf{k})$ is uniformly distributed over $\mathbf{k}$. This leads to an entropic uncertainty principle [1-3] such that:

$$H_\mathbf{q} + H_\mathbf{k} > 0$$

$$H_\mathbf{q} = -\sum_\mathbf{q} |\psi(\mathbf{q})|^2 \log\left(|\psi(\mathbf{q})|^2\right) \qquad H_\mathbf{k} = -\sum_\mathbf{k} |\phi(\mathbf{k})|^2 \log\left(|\phi(\mathbf{k})|^2\right) \qquad \text{S(4)}$$

where $H_\mathbf{q}$ and $H_\mathbf{k}$ are the information entropies for measuring into the quantum state and quantum functional respectively. Equation S(4) means for any qudit quantum object, the sum of the information entropies for measuring into the $\mathbf{q}$-representation and into the $\mathbf{k}$-representation can never be zero – thus there is always an uncertainty when measuring both observables. A tighter lower bound of $H_\mathbf{q} + H_\mathbf{k}$ should depend on both the qudit dimension $p$ and the number of qudits $n$. The entropic perspective in Equation S(4) may have potential applications in quantum encryption using quantum states as the ciphertexts [4], where maximizing $H_\mathbf{q} + H_\mathbf{k}$ or balancing $H_\mathbf{q}$ and $H_\mathbf{k}$ may increase the security of the quantum encryption method.

In the usual position-momentum $\mathbf{x}$ - $\mathbf{k}$ duality, the momentum operator $\hat{\mathbf{k}}$ is related to the translation operator $\hat{T}(\mathbf{x})$ by $\hat{T}(\mathbf{x}) = e^{-i\mathbf{x}\cdot\hat{\mathbf{k}}}$, and $\hat{\mathbf{k}} = -i\nabla$ is obtained as the infinitesimal generator of $\hat{T}(\mathbf{x})$. The $\mathbf{q}$ - $\mathbf{k}$ spaces are discrete so there can be neither derivative nor infinitesimal generator, but we recognize that on any single dimension of $\hat{\mathbf{k}}$, the observable $\hat{k}$ is still related to the translation operator $\hat{T}(q)$ by:

$$\hat{T}(q) = e^{-2\pi i \hat{k} \cdot q / p}, \qquad \hat{T}(q)|q_0\rangle = |q_0 + q\rangle \qquad \text{S(5)}$$

where $\hat{T}(q)$ translates any state $|q_0\rangle$ to $|q_0 + q\rangle$, similar to the usual translation $\hat{T}(x)|x_0\rangle = |x_0 + x\rangle$. In the $k$-representation the matrix of $\hat{T}(q)$ is $T_k(q) = diag(1, e^{-2\pi i q/p}, e^{-2\pi i \cdot 2q/p}, ..., e^{-2\pi i (p-1)q/p})$ and in Equation S(1) $K_k = diag(0,1,...,p-1)$, thus Equation S(5) is correct, and for multiple dimensions we also have $\hat{T}(\mathbf{q}) = e^{-2\pi i \hat{\mathbf{k}} \cdot \mathbf{q}/p}$. Therefore $\hat{\mathbf{k}}$ and $\hat{T}(\mathbf{q})$ are related by an exponential form, similar to the usual relation between the momentum $\hat{\mathbf{k}}$ and translation $\hat{T}(\mathbf{x})$.

## S2. The case of *d* not prime.

In all the discussions in the main text we have assumed the dimension of a single qudit *d* is a prime number *p*. Now when *d* is not a prime number, $\mathbb{Z}_d$ is not a field. Mathematically, a field requires that any element of the field except 0 has a multiplicative inverse. In $\mathbb{Z}_d$, if the non-prime *d* can be factored as $p_1 \cdot p_2$, then $p_1 \cdot p_2 = d \mod d = 0$ and thus neither $p_1$ nor $p_2$ can have an inverse. Now the consequence of $\mathbb{Z}_d$ not being a field is although we can still define $\mathbf{q} = (q_1, q_2, ..., q_n)$ and $\mathbf{k} = (k_1, k_2, ..., k_n) = k_1 q_1 \oplus k_2 q_2 \oplus ... \oplus k_n q_n$ with $q_j$ and $k_j$ taking integer values from 0 to $d-1$ and $\oplus$ being addition modulo $d$ – in the same way as Equations (1) and (2) in the main text – we can no longer consider **q** and **k** as vectors. This is because vector spaces must have scalars from a field. However, perhaps the most surprising thing is, other than **q** and **k** no longer being vectors, the entire discussion in the main text from Section 2.1 through Section 2.3 still holds when *d* is not a prime! This includes **k** being duals of **q**, **k** generating its own quantum space, the Fourier transform between **q** and **k**, the planewave interpretation of $|\mathbf{k}\rangle$, the partition interpretation of $|\mathbf{k}\rangle$, the physical realization of $|\mathbf{k}\rangle$ and $|\phi\rangle = \sum_\mathbf{k} b_\mathbf{k} |\mathbf{k}\rangle$, and the physical observable of $\hat{\mathbf{k}}$. The reason for this is the mathematical foundation of the entire discussion is the Pontryagin duality [5, 6] that can be defined on any abelian group (i.e. a group whose group operation is commutative): although $\mathbb{Z}_d$ is not a field, it is nonetheless an abelian group and thus all the discussions from Section 2.1 through Section 2.3 follow because of Pontryagin duality.

Is the difference between *d* being prime and non-prime purely mathematical, or rather it could lead to certain physical consequences? This can be an interesting future direction.

**Supplementary Information References:**